\def\etal{{et al.\thinspace}}

\def\spose#1{\hbox to 0pt{#1\hss}}
\def\approxlt{\mathrel{\spose{\lower 3pt\hbox{$\sim$}}
        \raise 2.0pt\hbox{$$<$$}}}
\def\approxgt{\mathrel{\spose{\lower 3pt\hbox{$\sim$}}
        \raise 2.0pt\hbox{$>$}}}

\def\multleft#1{\hbox to size{\vbox {\halign {\lft{##}\cr #1}}\hfill}\par}
\def\multright#1{\hbox to size{\vbox {\halign {\rt{##}\cr #1}}\hfill}\par}

\def\today{\ifcase\month\or January\or February\or March\or April\or May\or
      June\or July\or August\or September\or October\or November\or December\fi
      \space\number\day, \number\year}

\def\boxit#1{\vbox{\hrule\hbox{\vrule\kern3pt\vbox{\kern3pt
          #1 \kern3pt}\kern3pt\vrule}\hrule}}

\def\Mpc{{\rm\thinspace Mpc}}

\def\mic{{\rm\thinspace $\mu$m}}

\documentstyle[epsfig,psfig]{mn}
\begin{document}
\hsize=6truein

\title[INT WFS high redshift, z$>$4.4, quasars.]
{First results from the Isaac Newton Telescope Wide Angle Survey:
The z$>$5 quasar survey}
\author[Sharp R.G., McMahon R.G., Irwin M.J., Hodgkin S.T.]
{\parbox[]{6.5in}
{Sharp R.G., McMahon R.G., Irwin M.J., Hodgkin S.T.}\\
\it Institute of Astronomy, Madingley Road, Cambridge CB3 0HA, UK\\
Email: rgs, rgm, mike, sth@ast.cam.ac.uk
}

\date{Accepted ???. Received ???; in original form Latex version: 2001, Mar 05}

\maketitle

\begin{abstract}
We report the first results of an observational program designed to 
determine the luminosity density of high redshift quasars
(z$>$5 quasars) using deep multi-colour CCD data.
We report the discovery and spectra of 3 i$<$21.5 high redshift
(z$>$4.4) quasars, including one with z$>$5. At z=5.17, this is
the fourth highest redshift quasar currently published. 
Using these preliminary results we derive an estimate of the 
M$\rm_B$$<$$-$25.0 (M$\rm_{AB1450}$$<$$-$24.5) quasar space density in 
the redshift range 4.8$<$z$<$5.8 of 3.6$\pm$2.5$\times$$10^{-8}$\Mpc$^{-3}$.
When completed the survey will provide a firm constraint on
the contribution to the ionizing UV background in the redshift range
4.5$-$5.5 from quasars 
by determining the faint end slope
of the quasar luminosity function.
The survey uses imaging data 
taken with the 2.5m Isaac Newton Telescope as part of the Public Isaac
Newton Group Wide Field Survey (WFS).
  This initial
  sample of objects is taken from two fields of effective area
  $\sim$12.5deg$^2$ from the final $\sim$100deg$^2$.
\end{abstract}


\section{INTRODUCTION}

The identification and study of quasars at high redshift provides key
diagnostic information on the early Universe. Over the last few years
the relationship between high redshift quasars and local galaxies has
been given an impetus from the realization that quiescent black holes may
lie at the centre of all local bulges including the Milky Way (e.g.
Magorrian etal, 1998). The formation and evolution of massive black holes is
fast becoming a main stream topic in theories of galaxy formation
and evolution (Haehnelt, Natarajan and Rees \cite{Haehnelt}; 
Kauffmann and Haehnelt \cite{Kauffmann}). 

High redshift quasars are also an important contributor to the 
UV radiation field in the early Universe. Specifically 
the lack of an optically thick, Gunn-Peterson trough in the spectra of 
high redshift objects (e.g. Songaila, Hu, Cowie, McMahon \cite{Songaila1999}).
implies that the Universe is highly ionized at z$\sim$5.5.
Estimates by Madau\cite{Madau1999} indicate that if the ionization is due to
starlight the inferred star formation rate at z=5 must be comparable to
or greater than the observed value at z=3. 
The direct detection of galaxies at z$\sim$3 via
the Lyman Break technique and the detection of
measurable Lyman continuum photons that escape these
galaxies (Steidel, Pettini and Adelberger \cite{Steidel}) has raised a
question over whether quasars or galaxies are the origin
of the UV photons that re-ionized the Universe at z$>$6.
Recent theoretical work (Miralda-Escude, Haehnelt and Rees
\cite{Miralda-Escude}; Madau, Haardt and Rees \cite{Madau1999})
indicates that a Gunn-Peterson trough 
should be present by z$\sim$6. A major uncertainty
in all of this work is a measure of the ionizing background from
quasars. 
The current uncertainty in the contribution of quasars to 
the UV background is predominantly caused by the
unconstrained faint end of the luminosity function.

The resent discovery
of several z$>$5 quasars by the Sloan Digital Sky Survey
(SDSS Fan \etal \cite{fan4}) will constrain the bright 
end (i$<$20.0) of the high
redshift luminosity function. We are undertaking a survey
1-2 magnitudes fainter than the SDSS program with the aim
of constraining the space density and evolution of 
lower luminosity quasars where the majority of the
ionizing flux arises.
In this paper we present results from
the first $\sim$10deg$^2$ of our survey. 
Unless stated otherwise we use conventional Vega magnitudes and
H$_{0}=$50 km s$^{-1}$ Mpc$^{-1}$, q$_{0}$=0.5.

\section{Survey imaging data}

\begin{figure} 
        \epsfig{file=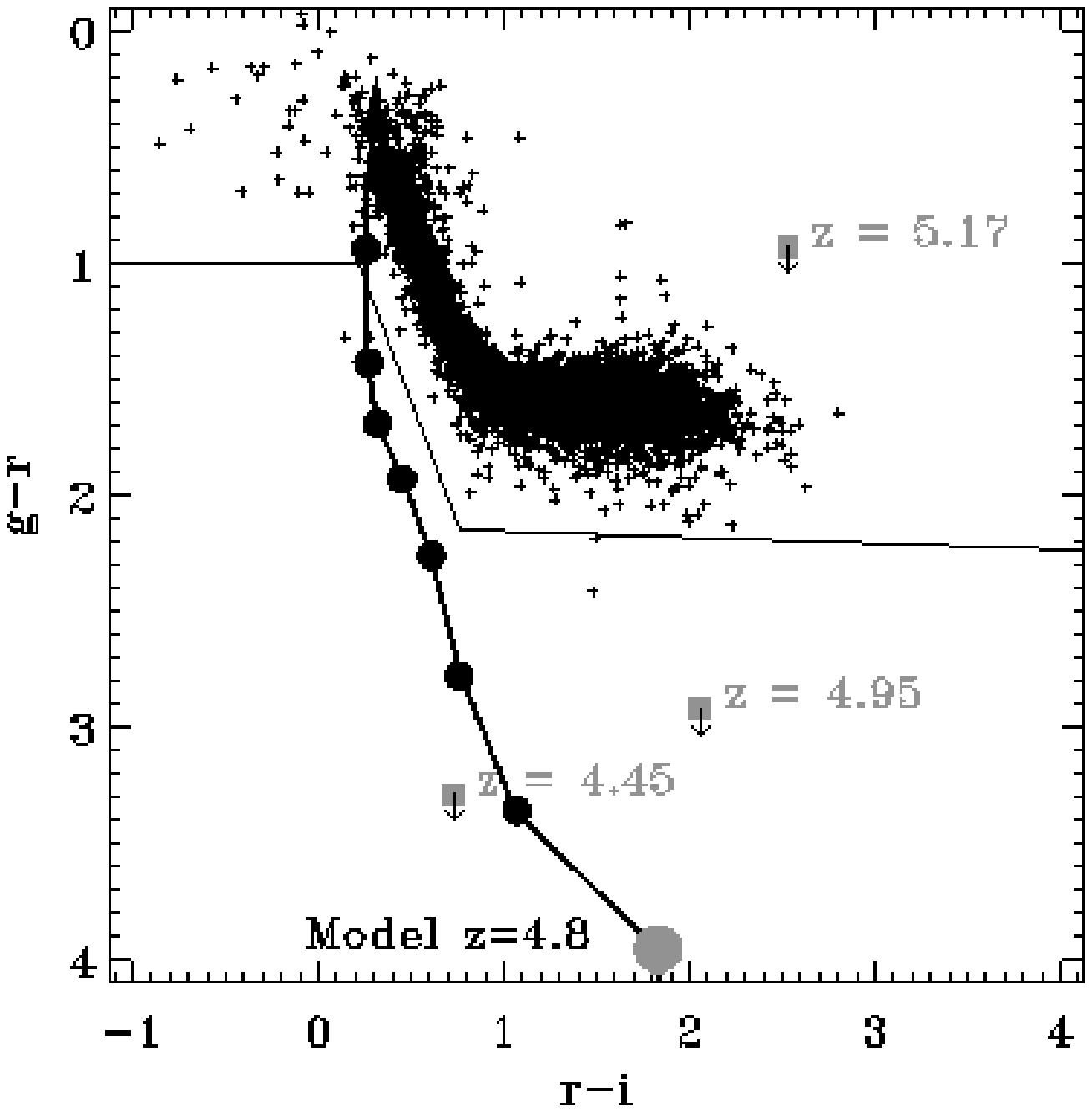,width=8cm}
        \epsfig{file=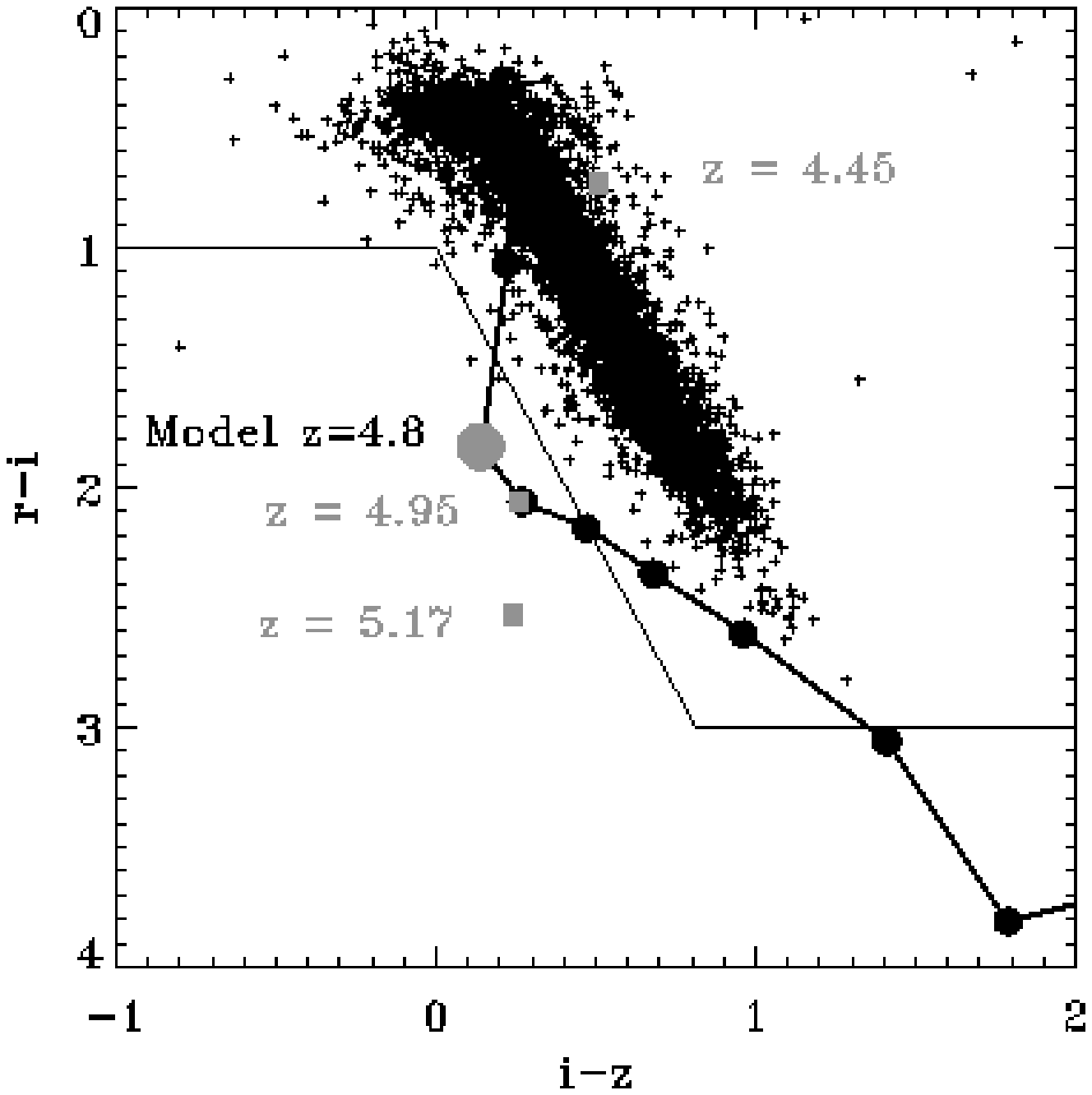,width=8cm}
 \caption[Colour-Colour plots]{\label{gri riz}The $\it{gri}$ and $\it{riz}$
two-colour diagrams.  Data from 8 pointings ($\sim$2deg$^2$) of the
INT WFS is show.  All objects are classified as stellar and have i$<$21.5
The newly discovered quasars (over plotted as squares) sit apart from
the stellar locus.  A model colour evolution track for quasars is overlaid.  Filled circles indicate a 0.2 step in redshift from the larger circle at a redshift of 4.8 in this model.}
\end{figure}

\begin{figure}
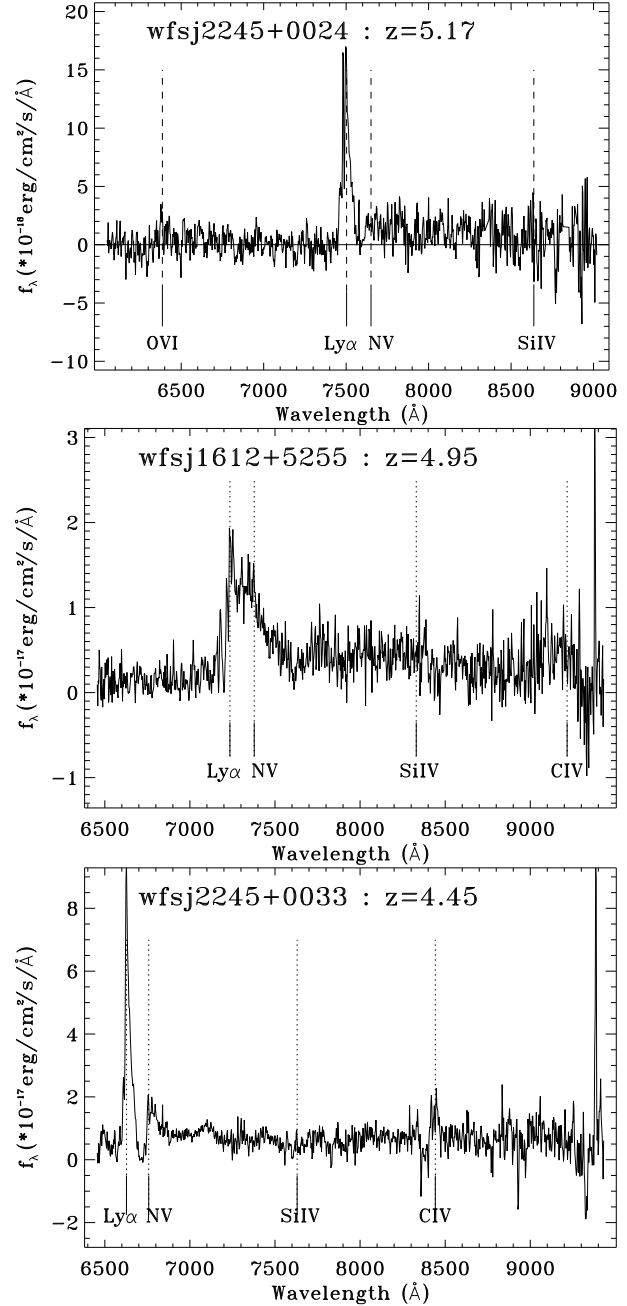
 
        \epsfig{file=qso5p15spec.epsi,width=8cm}
        \epsfig{file=qso4p93spec.epsi,width=8cm}
        \epsfig{file=qso4p44spec.epsi,width=8cm}
 \caption[Quasar spectra]{\label{qso spectra}Quasars are readily identified, even at low signal-to-noise, by the characteristic continuum break across the prominent Lyman-$\alpha$ emission line.  The identification of low level emission lines due to O\,{\sevensize{VI}}, Si\,{\sevensize{IV}} and C\,{\sevensize{IV}} at the correct wavelengths relative to Lyman-$\alpha$ support the redshift estimation for these objects.}
\end{figure}

We are using data from the Isaac Newton Telescope(INT) Wide Angle Survey (WAS)
(McMahon \etal \cite{McMahon2000}; http://www.ast.cam.ac.uk/$\sim$wfcsur/.
The survey is carried out with the prime focus 
Wide Field camera (WFC; Ives,Tulloch and Churchill \cite{Ives}) at
the 2.5m INT.
The WFC  consists of a close packed mosaic
of 4 thinned EEV42 2kx4k CCDs. 
with a pixel size of 13.5\mic\ corresponding to 0.33''/pixel
and effective field of view of 0.25deg$^2$.
The survey consists of 
single 
600secs
exposures in 5 wavebands ($\it{ugriz}$) 
over an area $\sim$100deg$^2$ to a nominal 5$\sigma$
limiting magnitude(Vega) of 23, 25, 24, 23 and 22 respectively.

The CCD mosaic data from all survey runs
is pipeline processed in Cambridge (Irwin and Lewis
\cite{IrwinLewis2000}) to produce
photometrically and astrometrically calibrated images
(rms(internal)=0.1''),
and  morphologically classified merged multicolour catalogues.
A preliminary photometric calibration has been applied to the
data to the $\pm$ 0.1 magnitude level.
For the current work we have only used data 
with seeing of $<$1.67 arcsec (FWHM of 5 pixels)
and stellar ellipticity (due to trailing) of better than $<$0.2
since this defines a practical
upper limit for reliable image morphological classification.
The current quasar candidate sample has a magnitude 
limit of $\it{i}$ or $\it{z}$$< \approx$21, the practical 
limit for acceptable 
signal-to-noise spectra with a 4m class telescope in
1800 seconds of observation.

At the time of the spectroscopic follow--up observations in June 2000, object
catalogues for data taken between August 1998 and
October 1999 were available.
This data is taken from two regions of the survey.  The first region covers
the ISO ELAIS N1 field at J1610+5430
(Oliver \etal \cite{Oliver}), the second region is in
the vicinity of at J2240+0000.
A total area of 12.5 deg$^2$ has been used.

\begin{table*}
\caption[Newly discovered quasars]{\label{new quasars}\footnotesize Photometry derived from the INT WFS.}
\begin{center}
\begin{tabular}{ccccccccccc}
 &  &  &  & \multicolumn{5}{c}{Vega system magnitudes}\\
Object & z &
\multicolumn{2}{c}{RA (J2000) Dec} 
&\ $\it{u}$  & $\it{g}$    & $\it{r}$   & $\it{i}$    &  $\it{z}$
& $\rm M_B$ &  M$\rm_{AB(1450)}$ \\
\hline
wfsj224524.2+002414 & 5.17 & 22 45 24.28  & +00 24 14.6 & \ $>$23.01  &\ $>$24.89 &\  23.96 &\ 21.43 &\ 21.19 & -24.9 & -24.3\\
wfsj161253.1+525543 & 4.95 & 16 12 53.11  & +52 55 43.5 & \ $>$23.00  &\ $>$25.12 &\  22.21 &\ 20.15 &\ 19.89 & -26.1 & -25.6\\
wfsj224531.0+003358 & 4.45 & 22 45 31.00  & +00 33 58.6 & \ $>$23.09  &\ $>$24.89 &\  21.60 &\ 20.87 &\ 20.36 & -25.5 & -25.0\\

\hline
\end{tabular}
\end{center}
\end{table*}

\section{Colour selection of candidates}

Figure \ref{gri riz} shows colour-colour plots for INT WFS data
in the $\it{griz}$ bands with the predicted quasar tracks as a function
of redshift overplotted. These colour tracks were computed assuming
an underlying quasar spectrum based on
a power law with spectral index ($S_\nu\propto \nu^{-0.5}$)
and with an emission line spectrum based on the composite spectrum
of Vanden Berk \etal \cite{vanden berk}.
The absorption model for the intervening Lyman-$\alpha$ forest is taken
from Madau(1995).  The stellar main sequence is
clearly visible as the heavily populated strip in the centre of the 
plots. 

Candidate high redshift objects are selected from the multi colour data
by selecting all objects that obey a set of colour criteria. 
To establish these criteria the stellar locus in the colour-colour
diagrams has been approximated using piecewise linear fits
in the colour space.
These fits are then used to establish the boundaries indicated
in Figure \ref{gri riz}.

Once selected by one of these colour criteria
all 5 ugriz images of an object are then visually examined
to check the validity of the photometry.  This step is required to
identify spurious sources such as objects in the diffraction spikes of
bright stars or satellites and asteroids that have moved significantly
between observations.
Candidate z$\geq$5 quasars are selected using the riz diagram
whereas 4$\leq$z$\leq$5 candidates
are selected from
the gri colour diagram.  At the current time the followup of the
lower redshift sample is incomplete due to limited spectroscopic time.

\begin{table*}
\begin{center}
\caption[]{\label{em line table}\footnotesize Emission line properties}
\begin{tabular}{ccccccccc}
Object & z & \multicolumn{4}{c}{Lyman-$\alpha$} & \multicolumn{3}{c}{C\,{\sevensize{IV}}} \\
       &   & Peak (z) & Cent.(z) & EW$_{obs}$ & EW$_{rest}$ & Cent.(z) & EW$_{obs}$ & EW$_{rest}$\\
\hline
wfsj2245$+$0024 & 5.17 & 7498\AA\ (5.17) & 7450\AA\ (5.16) & 594\AA\ & 115\AA\\
wfsj1612$+$5255 & 4.95 & 7237\AA\ (4.95) & 7318\AA\ (5.00) & 470\AA\ & 79\AA\  & 9112\AA\ (4.88)& 308\AA\ & 52\AA\ \\
wfsj2245$+$0033 & 4.45 & 6628\AA\ (4.45) & 6634\AA\ (4.46)& 276\AA\ & 51\AA\ & 8400\AA\ (4.42) & &\\
\hline
\end{tabular}
\end{center}
\end{table*}

\section{Spectroscopic observations}
Spectroscopic observations were obtained in two nights
(29th and 30th of June 2000) using the red arm of the
ISIS dual
beam spectrograph on the 4.2 meter William Hershel Telescope
on La Palma.
The R158R grating was used with the TeK4 CCD giving 
a spectral resolution of 2.90\AA\ per pixel and 
wavelength coverage of 2970\AA.  
Initially a central wavelength of 7900\AA\
was selected but during observations it became apparent
that a useful spectrum could not be taken of faint objects
longward of 9000\AA\ due to strong night sky line emission
and so a central wavelength of 7500\AA\ was
subsequently used.

The slit width was matched to the seeing
which varied over the course of the observing run in the 
range 0.8''--1.2''.
During the periods of poorer seeing($\geq$1'') brighter candidate objects 
were observed.  These generally proved to be low mass
late M and early L stars as expected.  
Conditions were photometric throughout the run and
observations of the spectrophotometric standards HZ 44, Ross 640 and
LDS 749B (ING identifiers sp1321+363, sp1626+368 and sp2129+000)
were used to flux calibrate the data.
Wavelength calibration was carried out using a mixture of
arc spectra and night sky lines.

Each observation was taken as a series of 900 second
exposures and coadded to improve cosmic ray rejection.
Total exposure times range from 900--2700 seconds.
Several candidate objects could be identified 
as late M and early L dwarf stars after only a single 900
second exposure.
Observation of these objects was truncated reducing the time
spent observing non-quasar candidates.

\section{Confirmed high redshift quasars}

The identification spectra of the confirmed high redshift (z$>$4.4)
quasars are shown in Figure \ref{qso spectra} and the emission line
properties are presented in Table \ref{em line table}.
Absolute B(Vega) band magnitudes and AB magnitudes at a rest-frame
wavelength of 1450\AA\ have been determined from the $\it{z}$ band
magnitudes assuming a power law spectral index of -0.5. We use
the $\it{z}$ band since it is unaffected by the Lyman-$\alpha$ forest or
any strong emission lines.

Redshift determination for high redshift objects is complicated since
in many spectra the only strong line visible is Lyman-$\alpha$(1215.7\AA).  
This 
is often blended on the red side with N\,{\sevensize{V}} emission(1240.1\AA) in quasar spectra
and has an asymmetric profile
due to absorption of the blue wing of the line by the Lyman-$\alpha$
forest. The onset of this absorption of the blue wing of the
line is quite abrupt and we take this as our redshift estimate for
the quasars. 


\smallskip
\indent{\bf wfsj224524.2+002414}
At z$=$5.17, this is currently the fourth highest redshift
quasar known.  This faint object is identified by the strong
Lyman-$\alpha$ line at 7498\AA\ and the break in the continuum
across the line due to Lyman-$\alpha$ forest absorption.
Lyman-$\alpha$ shows a highly irregular
profile with strong absorption features within the line.  A strong 
absorption feature is also at the expected wavelength 
of N\,{\sevensize{V}} emission redward of Lyman-$\alpha$.

\smallskip
\indent{\bf wfsj161253.1+525543}
This object shows a broad emission feature interpreted as
a blend of Lyman-$\alpha$ and N\,{\sevensize{V}}.  The wavelength of any associated C\,{\sevensize{IV}}
emission line is in a region of the spectrum heavily effected by emission
from the night sky making detection of an emission feature difficult.
At z=4.95 this quasar is the seventh highest redshift quasar known.

\smallskip
\indent{\bf wfsj224531.0+003358}
Strong Lyman-$\alpha$ emission dominates the spectra of this object.
Two clear absorption features, characteristic of the class of Broad
Absorption line (BAL) quasars, are seen associated with
N\,{\sevensize{V}} and C\,{\sevensize{IV}} emission lines. 

\smallskip
\par

The FIRST and NVSS radio surveys (Becker \etal \cite{becker};
Condon \etal \cite{condon}) cover our 
survey regions and no radio emission with $\rm S_{20cm}$$>$1mJy(5$\sigma$) is
detected for these quasars.
There is also no evidence for X-ray emission from ROSAT
All Sky Survey or PSPC pointed phase observations.

\begin{figure} 
        \epsfig{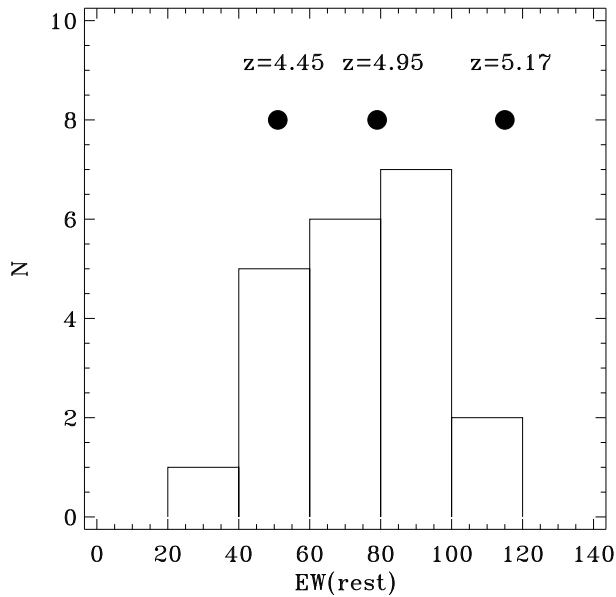}
\caption[Equivalent Widths]{\label{ew}The histogram of Lyman-$\alpha$ equivalent width for SDSS quasars with redshift z$>$4.4 is shown with the measurements from the three z$>$4.4 INT WFS quasars over plotted for comparison.}
\end{figure}

Figure \ref{ew} shows the equivalent width distribution of Lyman-$\alpha$
for the SDSS z$>$4.4 quasars compared with our new quasars.  
Despite
the strength of the line in our z=5.17 quasar it is comparable with
the strongest lined objects in the SDSS sample.  

\begin{figure} 
        \epsfig{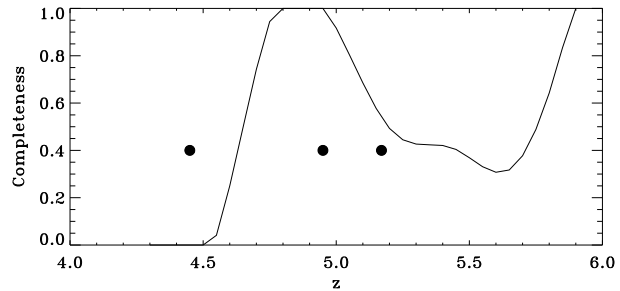}
\caption[Completeness riz]{\label{comp}An estimate of the survey
completeness is derived based on the model quasar spectrum described
in the text.  The fractional completeness is shown as a function of redshift
The confirmed quasars are marked as black circles.  At the current
limiting magnitude the survey is not sensitive to photometric errors
and so there is little change in completeness with magnitude.}

\end{figure}

\section{Discussion and conclusions}

The majority of the $\sim$250 z$>$4 quasars 
that are currently known (see http://www.ast.cam.ac.uk/\~{}quasars)
have been discovered
using photographic plates but
despite the relative ease with which z$>$4 quasars can be discovered 
(e.g. Storrie-Lombardi \etal \cite{Storrie-Lombardi} and refs therein)
for almost a decade the highest redshift object known was the 
optically selected quasar PC1247$+$3406 with z=4.89 (Schneider etal, 1991).  
The addition of the $\it{z}$ band to multi-colour surveys has changed this. 
The
first quasars with z$>$4.9 was discovered by Fan \etal \cite{fan99} using 
SDSS data in the griz bands. 
In this paper we report two new quasars at z$>$4.9
bringing the total sample of published quasars with redshift z$>$4.9 to nine,
of which 6 have z$\geq$5 (see Table \ref{known high}).

We use our sample to make the first estimate of the
space density of low luminosity quasars at z$\sim$5.
Figure \ref{comp} shows the completeness of our survey with redshift.
At the current magnitude limit for the spectroscopic sample the survey
is not sensitive to photometric errors and so there is no magnitude
effect on completeness. 
This has been computed
using the approach described in Fan \etal \cite{fan3} where
we use the Madau(1995) IGM model and an intrinsic
unabsorbed  quasar spectrum with a power law
continuum spectrum with a gaussian distribution of spectral indices
(mean=$-$0.5; dispersion($\sigma$=0.3) and an contribution
from emission lines.
The relative emission lines strengths 
are taken from the SDSS composite quasar spectrum
(Vanden Berk \etal \cite{vanden berk}) normalised
to a Lyman-$\alpha$ line strength distribution
(Rest frame Equivalent Width: Gaussian with mean 92.91 and sigma 24.0)
from Wilkes \etal(1986).
Using this completeness we calculate the effective
volume assuming H$_{0}=$50 km s$^{-1}$ Mpc$^{-1}$, q$_{0}$=0.5.

The INT WFS is ongoing and to date $\sim$10\% of the final survey 
imaging data
has been investigated for candidate quasars.  The final survey is expected
to contain 10--20 high redshift (z$>$5) low luminosity quasars
($\it{i}$$<$21.5 and should therefore constrain the luminosity density
to $\sim$20\%.
This statistically significant
sample of low luminosity quasars will allow the
determination of the faint end slope of the quasar luminosity
function and the break point in the distribution (analogue to L$^*$ for
galaxies).  This break point is required to allow estimation of the
contribution of quasars to the ionizing UV background at high redshift.
Our results, when combined with shallower surveys such as the SDSS, will
allow detailed studies of the AGN phenomenon at high redshift over a
range of UV luminosities. Our knowledge of the relationship between
accretion and the formation of black holes and galaxies will then 
not rely purely on the properties of the rare extreme luminosity quasars.

\begin{table}
\begin{center}
\caption[]{\label{known high}\footnotesize Known High redshift (z$>$4.9) quasars}
\begin{tabular}{ccccccccc}
Name             & z & \multicolumn{4}{c}{Mag} & Ref\\
                 &          & $\it{r}$       & $\it{i}$    & $\it{z}$  & M$\rm_{B}$ &\\
\hline
wfsj2245.6$+$0024 & 5.17  & 23.96    & 21.43 & 21.19 & -24.9      & 1\\
wfsj1612.7$+$5255 & 4.95  & 22.21    & 20.15 & 19.89 & -26.1      & 1\\
\hline
SDSS1044$-$0125    & 5.80 &$>$22.83  & 21.40 & 18.67 & -27.5      & 2\\
RD J0301$+$0020    & 5.50 & 26.2     & 23.38 & 22.84 & -23.4      & 3\\
SDSSp J1208$+$0010 & 5.27 & 22.66    & 20.37 & 20.17 & -26.0      & 4\\
SDSSp J1204$-$0021 & 5.03 & 20.72    & 18.89 & 18.56 & -27.5      & 5\\
SDSSp J0338$+$0021 & 5.00 & 21.61    & 19.54 & 19.19 & -26.8      & 6\\
SDSSp J1605$-$0112 & 4.92 & 22.42    & 19.36 & 19.37 & -26.6      & 5\\
SDSSp J0211$-$0009 & 4.90  & 21.97    & 19.51 & 19.26 & -26.7      & 6\\
\hline
\end{tabular}
\end{center}
Refs:
1. This work;
2. Fan \etal \cite{fanz=5.8};
3. Stern \etal\cite{stern};
4. Zheng \etal \cite{zheng};
5. Fan \etal\cite{fanII};
6. Fan \etal\cite{fan99};
\end{table}

\section*{ACKNOWLEDGMENTS}
RGS acknowledges the receipt of a PPARC studentship. RGM thanks the
Royal Society for support.
This paper uses data that was made publically available through 
the Isaac Newton Groups' Wide Field Camera Survey Programme. 
The Isaac Newton
Telescope is operated on the island of La Palma by the Isaac Newton
Group in the Spanish Observatorio del Roque de los Muchachos of
the Instituto de Astrofisica de Canarias.

\begin{figure}
\centering 
        \epsfig{file=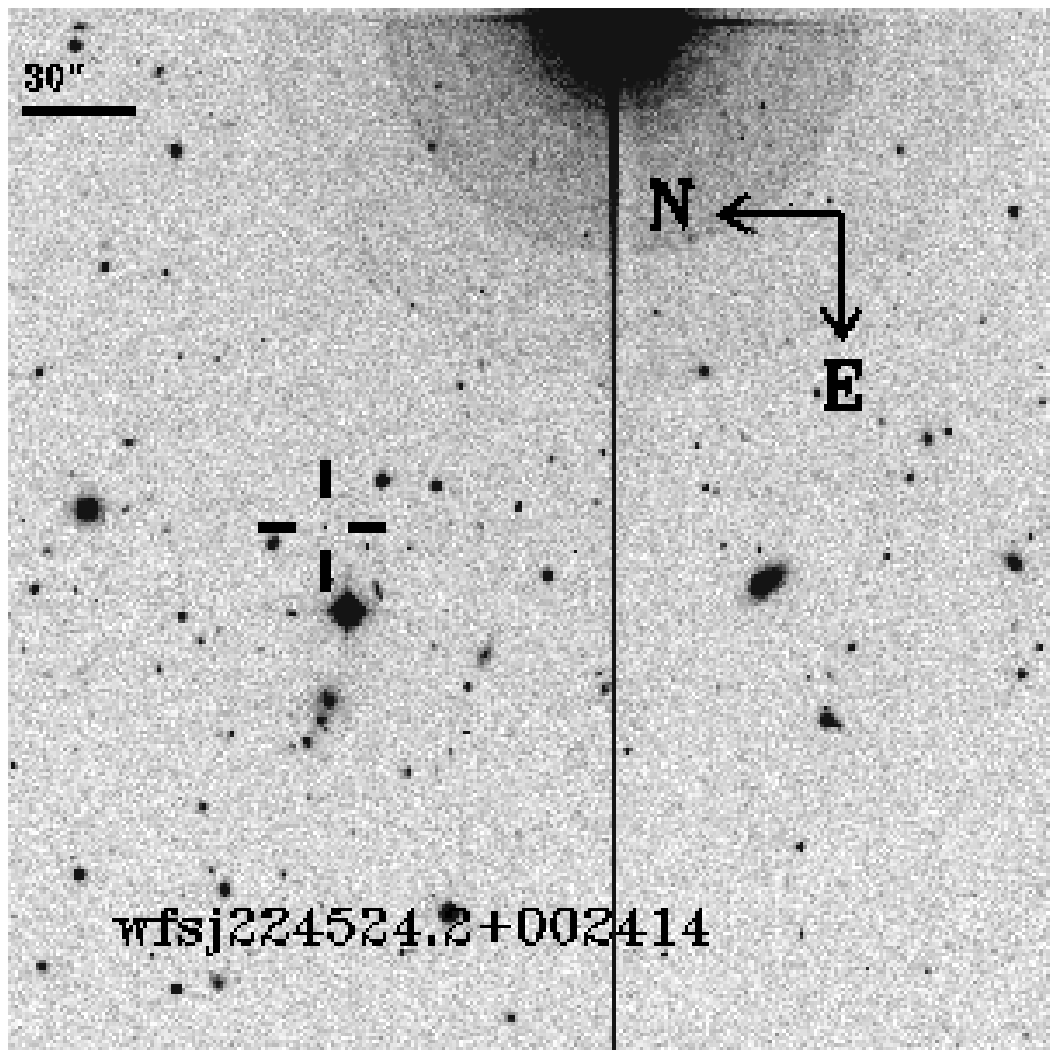,width=6cm}
        \epsfig{file=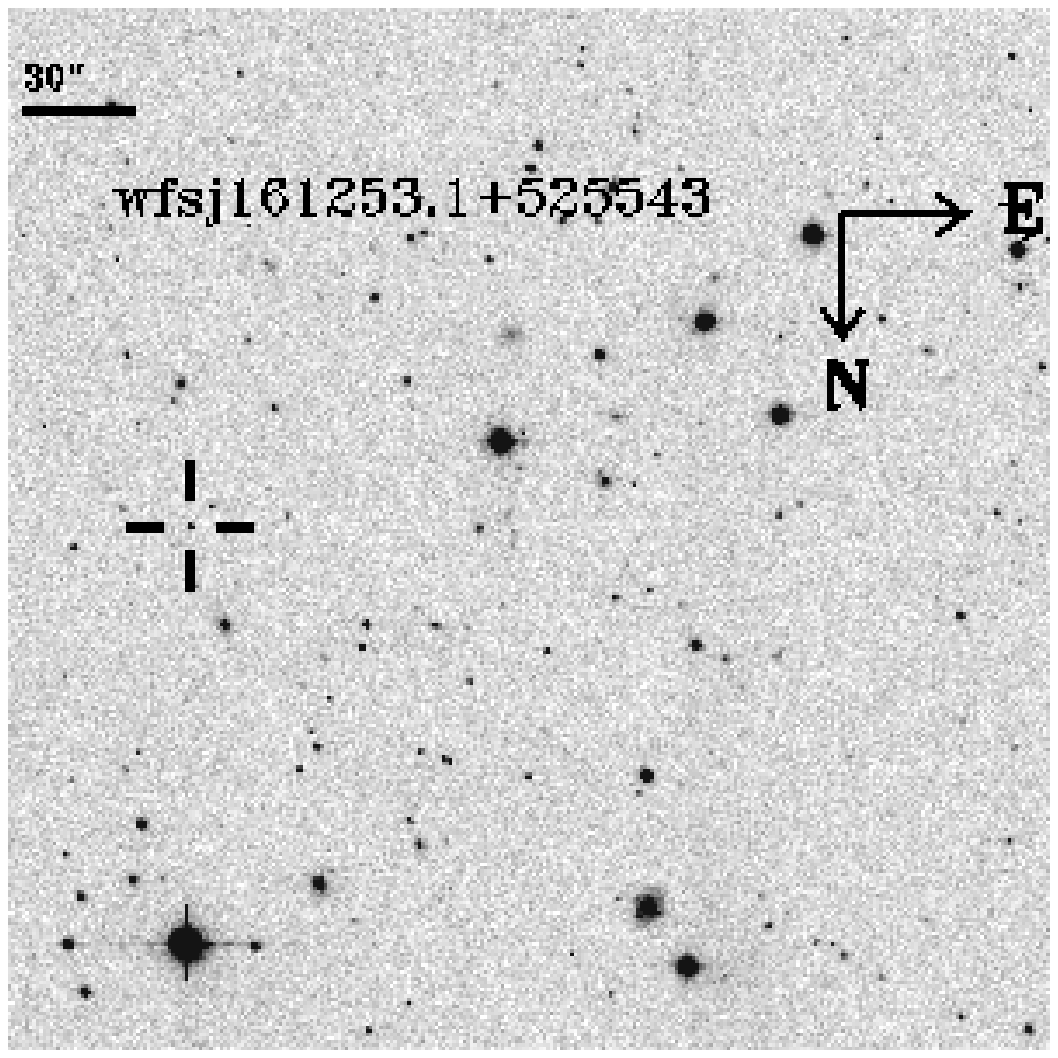,width=6cm}
        \epsfig{file=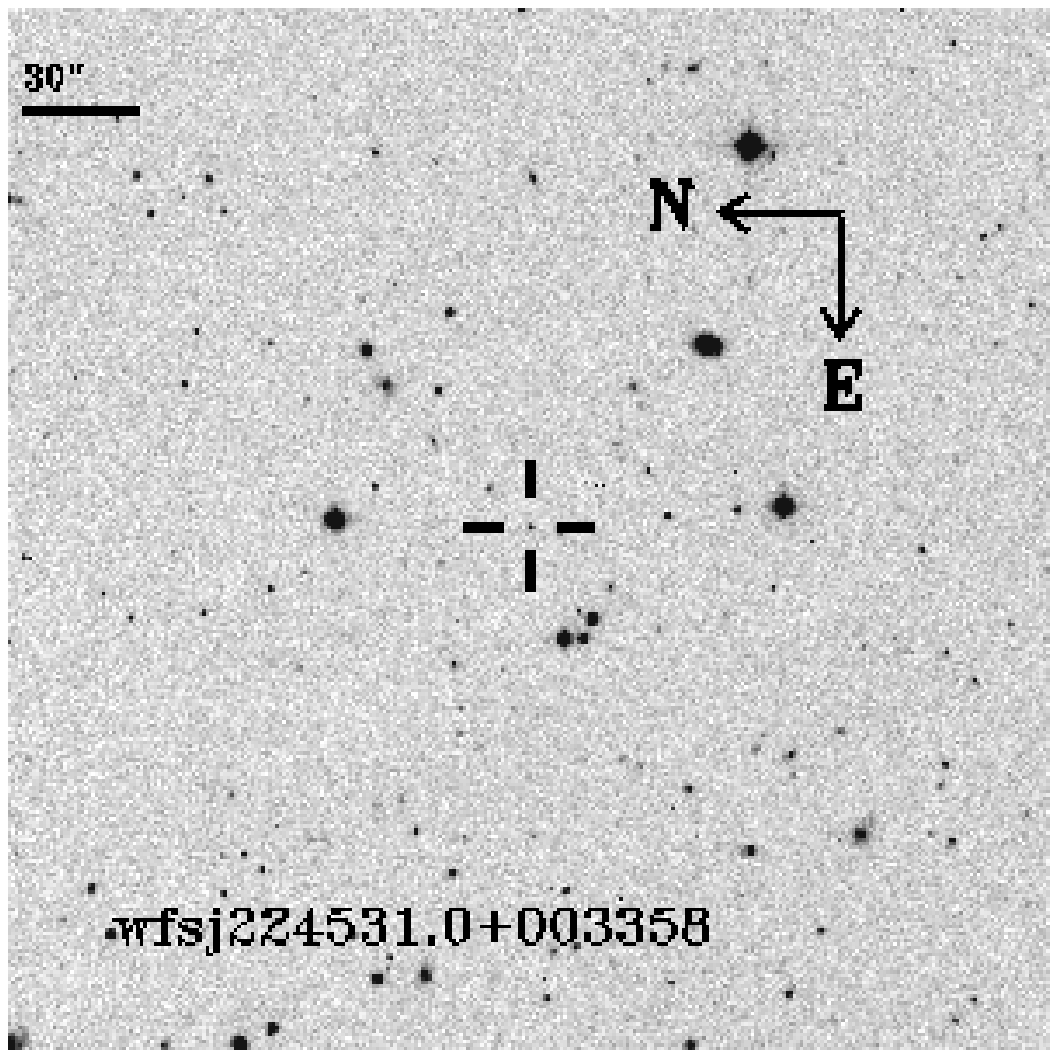,width=6cm}
\caption[fcharts]{\label{fcharts}Finding charts are taken from $\it{i}$ band images.  The charts are 4.4 arcminutes (800 pixels) square with the quasar marked by the cross hair.}
\end{figure}

\end{document}